\documentstyle[sprocl]{article}

\begin{document}

\title{RENORMALISING NN SCATTERING: IS POWER COUNTING POWERLESS?}
\author{Michael C. Birse}
\address{Theoretical Physics Group, Department of Physics and Astronomy,\\
University of Manchester, Manchester, M13 9PL, UK\\}
\maketitle

\abstracts{The renormalisation of NN scattering in theories with zero-range
interactions is examined using a cut-off regularisation where the cut-off is
taken to infinity, dimensional regularisation (DR) with minimal subtraction,
and DR with power-divergence subtraction. In the infinite cut-off limit power
counting breaks down: terms of different orders in the potential contribute to
the scattering amplitude at the same order. Minimal subtraction does yield a
systematic expansion, but with a very limited range of validity for systems
that have unnaturally large scattering lengths. For a finite cut-off, the
behaviour of the couplings as the cut-off is lowered shows that a theory with
a natural scattering length approaches an IR fixed point. In the corresponding
effective theory, loop corrections can be treated perturbatively. In contrast,
if there is an IR fixed point for systems with an infinite scattering length
it must be a nonperturbative one, with no power counting. For such systems,
power-divergence subtraction appears to yield a systematic expansion, but
with a different power counting from Weinberg's. However the scheme omits
IR divergent terms that would otherwise lead to nonperturbative behaviour
and so the interpretation of the fixed point remains unclear.}

\section{Introduction}

The possibility of applying the techniques of of effective field theory (EFT)
to nuclear physics was first raised by Weinberg~\cite{wein2} when he wrote down
power-counting rules for the low-momentum expansion of the {\it irreducible}
NN scattering amplitude. The raised the possibility of applying the techniques
of chiral perturbation theory (ChPT)~\cite{wein1,dgh,bkm} to nuclear forces.

By focussing on the NN potential (more precisely the two-nucleon irreducible
scattering amplitude) one avoids contributions where the two intermediate
nucleons are almost on-shell, and so have small energy denominators of order
${\cal O}(p^2/M)$ instead of ${\cal O}(p)$. However, this is the physics
responsible for nuclear binding, and so to describe nuclei with an EFT it is
not enough to write down a potential: one must be able to iterate it, by
solving a Schr\"odinger or Lippmann-Schwinger equation.

At this point one encounters a problem. The EFT is based on a Lagrangian with
local meson-nucleon couplings. As a result, multimeson exchange processes lead
to a potential that is highly singular at short distances. To renormalise these
short-distance (UV) divergences one has to introduce counterterms in the 
Lagrangian. These take the form of contact interactions such as
$(\overline\Psi\Psi)^2$ (at leading order in the momentum expansion) and
$(\overline\Psi\Psi)(\overline\Psi\nabla^2\Psi) +{\rm H.c.}$ (at
next-to-leading order). However, such interactions correspond to
$\delta$-function potentials, and the resulting scattering equations only make
sense after a further regularisation and renormalisation. Quite a few schemes 
have been explored recently for this$\,^{5-23}$.
The main question addressed by this workshop is: Can this renormalisation be
done while maintaining a useful and systematic organising scheme (power
counting) for the potential?

Here I am using systematic to imply that the coefficients appearing in the
potential should be meaningful beyond the particular calculation of NN
scattering at some order in a momentum expansion that has been used to fix
them. This means that higher-order terms in the potential should not contribute
to the scattering in the same way as lower-order ones~\cite{drp}, and so
changes in the coefficients should be small when includes higher-order terms
in the EFT. It also means that the values of these coefficients should be
applicable to calculations of other processes~\cite{tdc}.

In section 2, I introduce the EFT used to describe $s$-wave NN scattering at
very low energies. The use of a cut-off regulator is described in section 3,
along with the problems encountered if one tries to take the cut-off to
infinity. Section 4 gives more details of the behaviour if the cut-off is left
finite. Approaches based on dimensional regularisation are described in
sections 5 (the minimal subtraction scheme) and 6 (power divergence
subtraction). In section 7 I compare the results obtained in all these
approaches. Apart from the extended discussion of finite cut-offs, these
sections are essentially the talk presented at the workshop. Finally, in
section 8, I examine the the finite cut-off and power divergence subtraction
schemes from a renormalisation-group viewpoint. These ideas, which were 
developed following discussions at the workshop, suggest that the low-energy
EFT of systems with natural scattering lengths corresponds to a perturbative
IR fixed point. In contrast, at least for cut-off regularisation, the fixed
point for systems with infinite scattering lengths seems to be a
nonperturbative one, with no power counting.

\section{The model}

Like most of the other contributors to this session, I consider an EFT 
where all mesons (including pions) have been integrated out. Although such a 
theory is only relevant to extremely low-energy NN scattering, it can be used 
to address questions of principle. The basic potential in this model contains
contact interactions only. To second order in the momentum expansion it has 
the form
\begin{equation} \label{eq:pot1}
V(k',k;E)=C_0+C_2(k^2+k'^2)+\cdots,
\end{equation}
where only the terms relevant to $s$-wave scattering have been included.

In (\ref{eq:pot1}) I have allowed for a possible energy dependence of the
potential. This is because, as described below, renormalisation of the
scattering equation naturally leads to counterterms proportional to powers of
the energy. Energy dependence of the potential should not be too surprising
since it can arise whenever degrees of freedom are eliminated from a
Schr\"odinger equation~\cite{fesh}. Indeed energy-dependent contact terms are
naturally required to renormalise, for example, contributions to the two-pion
exchange potential. Such terms have not normally been considered in EFT
treatments of NN scattering, where energy dependence is usually eliminated
using the equation of motion. Since energy- and momentum-dependent terms in
the potential behave differently off-shell, they can give different results
when iterated in a Lippmann-Schwinger equation. For example, with a cut-off
regulator that is taken to infinity, the bare couplings are renormalised
differently~\cite{rbm}. Hence, until a consistent power-counting scheme has
been established, energy-dependent terms ought to be included in the potential.

In treating the scattering non-perturbatively, it is convenient to work
with the reactance matrix~\cite{newton}, $K$, rather than the scattering
matrix, $T$. The off-shell $K$-matrix for $s$-wave scattering satisfies a
Lippmann-Schwinger equation that is very similar to that for $T$:
\begin{equation} \label{eq:lse}
K(k',k;E)=V(k',k;E)
+\frac{M}{2\pi^2}{\cal{P}}\int_{0}^{\infty}q^2dq\,\frac{V(k',q;E)K(q,k;E)}
{{p^2}-{q^2}}.
\end{equation}
In this expression and throughout this paper, $p=\sqrt{M E}$ denotes the
on-shell value of the relative momentum. Note that the definition of $K$ here
differs from the more standard one~\cite{newton} by a factor of $-\pi$. The
equation is similar to the one for the $T$-matrix except that the Green's
function satisfies standing-wave boundary conditions. This means that the usual
$i\epsilon$ prescription for the integral over $q$ is replaced by the
principal value (denoted by ${\cal P}$). As a result, the $K$ matrix is real
below the threshold for meson production.

The inverse of the on-shell $K$-matrix differs from that of the on-shell
$T$-matrix by a term $iM p/4\pi$, which ensures that $T$ is unitary if $K$ is
Hermitian. This allows the effective-range expansion~\cite{bj,bethe,bw} to be 
written as an expansion of $1/K$:
\begin{eqnarray} \label{eq:ere}
\frac{1}{K(p,p;E)}&=&\frac{1}{T(p,p;E)}-{iMp\over 4\pi}\\
&=&-{M\over 4\pi}p\cot\delta(p)\nonumber\\
&=&-\frac{M}{4\pi}\left(-\frac{1}{a}+\frac{1}{2}r_{e}p^2
+\cdots\right),\nonumber
\end{eqnarray}
where $a$ is the scattering length and $r_e$ is the effective range.

\section{Cut-off regularisation}

To regulate the divergences associated with the $\delta$-function potential
(\ref{eq:pot1}), one can smear it out over distances of the order of
$1/\Lambda$ by introducing a separable form factor~\cite{rbm}:
\begin{equation}\label{eq:potreg}
V(k',k;E)=f(k'/\Lambda)\left[C_0+C_2\,(k^2+k'^2)\right]f(k/\Lambda),
\end{equation}
where the form factor $f(k/\Lambda)$ satisfies $f(0)=1$ and falls off 
rapidly for momenta above the cut-off scale $\Lambda$.

The resulting potential has a two-term separable form and so the corresponding
Lippmann-Schwinger equation can be solved using standard
techn\-iques~\cite{newton}. The off-shell $K$-matrix obtained in this way is
\begin{eqnarray} \label{eq:koff}
&&K(k',k;E)=f(k'/\Lambda)f(k/\Lambda)\\
&&\quad\qquad\times\frac{1+\frac{C_2}{C_0}(k^2+k'^2)
+\frac{C_2^2}{C_0}\left[I_2(E)-(k^2+k'^2)I_1(E)
+k^2k'^2I_0(E)\right]}{\frac{1}{C_0}-I_0(E)-2\frac{C_2}{C_0}I_1(E)
-\frac{C_2^2}{C_0}\left[I_2(E)I_0(E)-I_1(E)^2\right]},\nonumber
\end{eqnarray}
where the integrals $I_n(E)$ are given by
\begin{equation}\label{eq:qint}
I_n(E)=\frac{M}{2\pi^2}{\cal{P}}\int_{0}^{\infty}\frac{q^{2n+2}f^2
(q/\Lambda)}{{p^2}-{q^2}}dq,
\end{equation}
with $p^2=ME$ again. The corresponding expression for $T$ has also been
obtained by the Maryland group~\cite{md4,md5}, who also considered regulating
the theory by simply cutting off the momentum integrals.\footnote{A similar 
result has also been obtained by Vall {\it et al.}~\cite{vklst}, in a rather
different notation.} That procedure leads to a similar expression to
(\ref{eq:koff}) but without the factors of $f(k/\Lambda)$ outside.

By expanding the integrals (\ref{eq:qint}) in powers of the energy (or $p^2$),
one can extract their divergent parts:
\begin{equation} \label{eq:expand}
\frac{I_n(E)}{M}=-\sum_{m=0}^{n}A_m\Lambda^{2m+1}p^{2(n-m)}
+\frac{F(p/\Lambda)}{\Lambda}p^{2(n+1)},
\end{equation}
where the dimensionless integrals $A_m>0$ and $F(p/\Lambda)$ are finite as
$\Lambda\rightarrow\infty$.\footnote{The integral $A_m$ corresponds, up to a
factor of $-M\Lambda^{2m+1}$, to $I_{2m+1}$ in the notation of
Refs.~\cite{md4,md5}.} Note that, as mentioned above, the divergences appear
multiplying powers of $E\,(=p^2/M)$ and so it is natural to introduce
energy-dependent counterterms to cancel them.

Consider first the limit where $\Lambda\rightarrow\infty$. In this limit the 
outside factors of $f(k/\Lambda)$ in $K$ can be replaced by unity and the final
term in (\ref{eq:expand}) vanishes. The on-shell $K$-matrix is then given by
\begin{equation}\label{eq:kon}
K(p,p;E)={N(p)\over N(p)A_0M\Lambda+(1+C_2A_1M\Lambda^3)^2},
\end{equation}
where the numerator function is
\begin{equation}\label{eq:num}
N(p)=C_0-C_2^2A_2M\Lambda^5+p^2C_2(2+C_2A_1M\Lambda^3).
\end{equation}
For this to remain finite as $\Lambda\rightarrow\infty$, the coefficients
must vanish like
\begin{equation} \label{eq:scale}
C_0\sim \frac{1}{M\Lambda},\qquad\qquad C_2\sim\frac{1}{M\Lambda^3}.
\end{equation}
This scale dependence of $C_0$ is that same as that found by
Weinberg~\cite{wein2} and Adhikari and coworkers~\cite{afg}.

The leading terms in both the numerator and the denominator of (\ref{eq:kon})
cancel, and so a finite result is obtained from subleading pieces of $C_{0,2}$.
If one assumes that $C_{0,2}$ depend analytically on the cut-off~\cite{rbm},
\begin{equation}\label{eq:subl}
C_0={\alpha_0\over M\Lambda} +{\beta_0\over\Lambda^2},\qquad\qquad
C_2={\alpha_2\over M\Lambda^3}+{\beta_2\over\Lambda^4},
\end{equation}
then for $\Lambda\rightarrow\infty$ the off-shell $K$-Matrix is
\begin{equation}\label{eq:result}
K(k',k;E)={1\over M^2}{\alpha_0-\alpha_2^2A_2\over
\beta_0A_0+2\beta_2\bigl[A_1+\alpha_2(A_1^2-A_0A_2)\bigr]},
\end{equation}
where $\alpha_{0,2}$ satisfy the ``fine-tuning" condition
\begin{equation}
\alpha_0A_0-\alpha_2^2A_0A_2+(1+\alpha_2A_1)^2.
\end{equation}
Although this result gives a finite scattering length, it has no energy or
momentum dependence, and so the effective range is zero. 

Therefore, under the assumption of analytic dependence on the cut-off, a
non-zero effective range can only be obtained as the cut-off is removed if the
coefficients in the potential are allowed to depend on energy. Either or both
of the subleading coefficients $\beta_{0,2}$ may be given a linear energy
dependence to generate a finite scattering length and effective range. Note
that an energy-dependent $\beta_0$ leads to the energy appearing the potential
with a coefficient of order $\Lambda^{-2}$ while the leading coefficient of
the momentum-dependence ($C_2$) is of order $\Lambda^{-3}$. This shows that,
in the absence of systematic power counting, energy and momentum dependence
need not be equivalent.

The Maryland group~\cite{md4,md5} have taken the on-shell $K$-matrix
(\ref{eq:kon}) and treated it somewhat differently, by demanding that it match
the observed scattering length and effective range for any $\Lambda$. The 
results can be expanded as a power series in $\Lambda^{-1/2}$, with the same
leading terms as above (\ref{eq:scale}). Terms up to order $\Lambda^{-2}$
beyond the leading order must be kept to obtain a finite scattering length.
Although a finite effective range can be obtained without introducing energy
dependence into the potential, this effective range cannot be
positive~\cite{md1,md2}, as required by Wigner's bound on the momentum
dependence of phase shifts~\cite{wigner}. This means that energy dependence of
the $C$'s is required to obtain a positive effective range. (A positive
scattering length can also be obtained by letting the potential become
complex~\cite{vank}. However this will lead to violations of unitarity for any
finite $\Lambda$.)

In either case (analytic or nonanalytic dependence on $\Lambda$) one finds
that terms of different orders in the potential are contributing at the same
order in the expansion of $K$. Indeed the bare parameters are not uniquely
determined, if one demands matching to scattering observables only as
$\Lambda\rightarrow\infty$~\cite{rbm}. Weinberg's power counting has therefore
broken down for cut-off regularisation with $\Lambda\rightarrow\infty$.

\section{Finite cut-off}

If taking the cut-off to infinity destroys power counting, one might hope that
keeping it finite could avoid the problem~\cite{lep}. It does so, but only in
natural systems, where $a\sim r_e$. In such cases one can choose a cut-off
that is well below the scale of the omitted physics, $\Lambda<\!\!<1/r_e$,
without needing any fine tuning to get the scattering length~\cite{md5}.

If the cut-off is taken to be at or below the scale $r_e$ of the omitted
physics, one can no longer omit the terms involving inverse powers of
$\Lambda$ in (\ref{eq:expand}). To order $p^2$ one has to keep the leading
term in the expansion of $F(p/\Lambda)$,
\begin{equation}\label{eq:fexp}
F(p/\Lambda)=-B_1+{\cal O}(p^2).
\end{equation}
The $K$ matrix can then be obtained from (\ref{eq:kon}) by the substitution
$A_0\rightarrow A_0+B_1p^2/\Lambda^2$, where $B_1$ is another dimensionless
integral. The corresponding effective range expansion can be written
\newpage
\begin{eqnarray}\label{eq:erefc}
\frac{1}{K(p,p;E)}&=&M\Lambda\left[A_0
+{(1+\hat C_2A_1)^2\over \hat C_0-\hat C_2^2A_2}\right]\\
&&+{Mp^2\over\Lambda}\left[B_1
-\left({1+\hat C_2A_1\over \hat C_0-\hat C_2^2A_2}\right)^2
(2+\hat C_2A_1)\hat C_2\right]+\cdots,\nonumber
\end{eqnarray}
in terms of the dimensionless couplings
\begin{equation}\label{eq:dless}
\hat C_0=M\Lambda C_0,\qquad\qquad \hat C_2=M\Lambda^3 C_2.
\end{equation}

For $\Lambda<\!\!<1/a\sim 1/r_e$, the coefficients $C_{0,2}$ can be expanded
in powers of $\Lambda$, with the leading behaviours~\cite{md5}
\begin{equation} \label{eq:natscale}
C_0\sim \frac{r_e}{M}+{\cal O}(\Lambda^{-1}),\qquad\qquad 
C_2\sim\frac{r_e^2}{M\Lambda}+{\cal O}(\Lambda^0),
\end{equation}
instead of (\ref{eq:scale}). Note that the higher-order terms in these
coefficients are suppressed by positive powers of $\Lambda$, in contrast to
the expansion for large cut-offs discussed in the previous section. Loop
effects are suppressed by powers of $r_e\Lambda$, and a systematic
organisation of the calculation is possible~\cite{md5}. Note that the
$1/\Lambda$ behaviour of $C_2$ is needed to cancel the contribution
$B_1/\Lambda$ to the effective range in (\ref{eq:erefc}). In a natural theory
such contributions can be cancelled without spoiling the power counting.

The situation is quite different in systems with bound states close to
threshold, such as $s$-wave NN scattering. In these the scattering length is
unnaturally large, $a>\!\!>r_e$. If one chooses a cut-off $\Lambda$ that is
much larger than $1/a$, but still well below $1/r_e$, then one has again to be
careful to keep the piece of order $1/\Lambda$ in the effective range term of
(\ref{eq:erefc}). This $1/\Lambda$ piece is much larger than the effective
range and so must be cancelled by a similar piece in $C_2$, which leads to
$C_{0,2}$ having the same dominant behaviour as in (\ref{eq:scale}). However
these cannot be regarded as the leading terms in power series in either
$\Lambda$ or $1/\Lambda$, since there are corrections involving powers of
$1/\Lambda a$ as well as $\Lambda r_e$. Also, as in the case of a very large
cut-off, $C_2$ contributes to the scattering length at order unity. Hence, in
systems with unnaturally large scattering lengths, there is no systematic
power counting. Nonetheless this approach may still be useful as a tool for
analysing low-energy NN scattering without introducing too many
parameters~\cite{orvk,md3,pkmr,sf}.

\newpage
\section{Minimal subtraction}

The predictions of a quantum field theory should be independent of the 
regulator, yet dimensional regularisation (DR) seems to yield quite different
results, evading the problems just discussed~\cite{ksw}. In the minimal
subtraction scheme, DR detects only logarithmic divergences., which show up
as poles at $D=4$ dimensions. The loop integrals $I_n(E)$ introduced in
(\ref{eq:qint}) above contain only power-law divergences, and so minimal
subtraction sets them to be identically zero. As a result the $K$-matrix is 
given by the first Born approximation,
\begin{equation}\label{eq:born}
K(k',k;E)=V(k',k;E).
\end{equation}

The problem with this scheme is that bound states close to threshold lead to
the on-shell $K$-matrix varying rapidly with energy. In such cases, which the
scattering length is unnaturally large, and and the low-momentum expansion of
$K$, and hence also that of the potential, is only valid for
$p<\sqrt{2/ar_e}$.~\cite{ksw,lm} The minimal subtraction scheme, while
systematic, is hardly useful in the case of interest, $s$-wave NN scattering.

\section{Power divergence subtraction}

Recently Kaplan, Savage and Wise~\cite{ksw2,dbk} (see also Ref.~\cite{geg})
have suggested an alternative renormalisation scheme that might allow one to
do better using DR. When continued to $D$ space-time dimensions, the loop
integrals (\ref{eq:qint}) take the form
\begin{eqnarray}\label{eq:drloop}
I_n(E)&\!\!=\!\!&{M\over(2\pi)^{D-1}}\,\left({\mu\over 2}\right)^{4-D}
{\cal P}\!\int\!{q^{2n}\over p^2-q^2}d^{D-1}\!{\bf q}\\
&\!\!=\!\!&-{M\over(2\sqrt\pi)^{D-1}}\,{(\mu/2)^{4-D}\over \Gamma\left(
{D-1\over2}\right)}{\cal P}\!\int_0^\infty\! dx\,{x^{(D+2n-3)/2}\over x-p^2}
\nonumber\\
&\!\!=\!\!&-{Mp^{2n}\over(2\sqrt\pi)^{D-1}}\,{(\mu/2)^{4-D}\over \Gamma(
{D-1\over2})}\,{\rm Re}\!\left[(-p^2)^{(D-3)/2}\right]
\Gamma\left(\textstyle{D+2n-1\over 2}\right)
\Gamma\left(\textstyle{3-2n-D\over 2}\right).\nonumber
\end{eqnarray}
The final $\Gamma$-function in this expression has a pole at $D=3$ for any
$n$. This is the signal of a logarithmic divergence in three dimensions, or a
linear one in four. The power-divergence subtraction (PDS) scheme~\cite{ksw2}
keeps this piece, cancelling it against a counterterm with the same pole at
$D=3$ to leave
\begin{equation}\label{eq:pdsint}
I_n(E)=-A_0M\mu p^{2n},
\end{equation}
where $A_0=1/4\pi$.

The resulting on-shell $K$-matrix is 
\begin{equation}\label{eq:konpds}
K(p,p;E)=\left[{1\over C_0+2p^2C_2}+A_0M\mu\right]^{-1},
\end{equation}
and the corresponding scattering length is given by
\begin{equation}\label{eq:apds}
{1\over a}={4\pi\over M}\left({1\over C_0}+A_0M\mu\right).
\end{equation}
This shows that PDS contains the ``strength-range" cancellation needed to give
a large scattering length without requiring $C_0$ to be unnaturally small.
As a result one can choose $\mu>\!\!>1/a$. In this scheme, the scale
dependences of $C_{0,2}$ are
\begin{equation} \label{eq:pdsscale}
C_0\sim \frac{1}{M\mu},\qquad\qquad C_2\sim\frac{r_e}{M\mu^2}.
\end{equation}
If one chooses the scale $\mu$ to be of the same order as the momenta of
interest, $\mu\sim p$, and much less than the scale of the new physics,
$\mu<\!\!<1/r_e$, then $C_0$ must be treated to all orders but $C_2$ gives 
corrections that are suppressed by $pr_e$. Higher terms in the potential are
similarly suppressed by powers of $pr_e$ and so a systematic power counting
does exist in the PDS scheme, although it is not the one suggested by Weinberg.

Moreover the linearly divergent terms are ``universal" in the sense that they
have the same coefficient in all of the loop integrals, up to powers of $p^2$.
Powers of energy multiplying the integrals and powers of momentum appearing
inside the $I_n(E)$ both contribute in the same way, and so there is no
distinction between energy and momentum dependence of the potential, as
expected for a scheme with a systematic power counting. As noted by
Gegelia~\cite{geg} the same results can also be obtained by keeping the linear
divergences and carrying out a momentum subtraction at the unphysical point
$p=i\mu$.

PDS as described in Refs.~\cite{ksw2} keeps only subtraction terms arising
 from the linear divergences in the $I_n(E)$. One might ask whether subtraction
terms for the higher power-law divergences spoil the power counting. The
divergent integral over $x=q^2$ in (\ref{eq:drloop}) gives a $\Gamma$-function
with poles at $D=3,\ 1,\ ,\dots,\ 3-2n$ dimensions, corresponding to all the
divergences seen with a cut-off regulator in (\ref{eq:expand}). It happens that
all except the pole at $D=3$ are cancelled by zeros of factor
$\Gamma({D-1\over 2})^{-1}$ arising from the angular integral. This
cancellation would appear to be an artefact of continuing to numbers of spatial
dimensions of zero or less, where the angular integration no longer makes
sense. If one modifies DR by analytically continuing only the $q^2$ integral
then all the power-law divergences can be identified and subtracted to leave
\begin{equation} \label{eq:pdsexpand}
\frac{I_n(E)}{M}=-\sum_{m=0}^{n}A_m\mu^{2m+1}p^{2(n-m)}.
\end{equation}

Implemented in this way, PDS leads to a result that looks very like the one
obtained above using a cut-off (\ref{eq:expand}), except that the cut-off
$\Lambda$ has been replaced by $\mu$. This is a rather natural result if one
regards the scale $\mu$ introduced by DR as a resolution scale. The requirement
that physics be independent of $\mu$ (renormalisation-group invariance) can
then be implemented by solving the equations for $C_{0,2}$ in terms of $a$
and $r_e$, exactly as in Ref.~\cite{md4,md5} for the cut-off case. There is
however one important difference between DR (\ref{eq:pdsexpand}) and the
cut-off (\ref{eq:expand}):\footnote{I am grateful to D. Phillips for pointing
this out.} the final UV finite, but IR divergent term is absent from the PDS
expression. This means that, with $\mu<\!\!<1/r_e$, one is not forced into
having coefficients with the scale dependence (\ref{eq:scale}). Instead the
$C_{0,2}$ continue to have the leading scale dependence (\ref{eq:pdsscale}).

In this modified version of PDS, the fitted value of $C_0$ does shift when
$C_2$ is included in the potential. However this shift is suppressed by $\mu
r_e$, and so is small for $\mu<\!\!<1/r_e$. The power counting of
Ref.~\cite{ksw2} therefore survives the inclusion of higher power-law
divergences.

\section{Discussion}

In the simplified model for $s$-wave NN scattering by short-range potentials,
taking the cut-off to infinity leads to a breakdown~\cite{md4,md5,rbm} of the
power counting proposed by Weinberg~\cite{wein2}. In the absence of a
consistent power counting, energy- and momentum-dependent terms in the
potential are not equivalent. In particular, as $\Lambda\rightarrow\infty$,
energy dependence is essential to get the correct effective range.

This is unsurprising if one interprets the (renormalised) short-range
potential as simply imposing a boundary condition on the logarithmic
derivative of the wave function, as suggested in Manchester some time
ago~\cite{bp}. In the $\Lambda\rightarrow\infty$ limit, this boundary
condition is imposed at the origin~\cite{bp,breit,bf,shir,do,few}. Since the
logarithmic derivative of the wave function at the origin is just
$p\cot\delta(p)$ (cf.~the effective range expansion (\ref{eq:ere})), this
implies a zero effective range, unless the boundary condition depends on
energy~\cite{vank}.

One alternative is not to take the cut-off to infinity, but to set it to some
scale roughly corresponding to the physics that has been omitted from the
effective theory, checking that results do not depend too strongly on the
precise value of the cut-off~\cite{lep}. This form of regularisation is
appealing, since we know that the physics described by the contact
interactions is not truly zero-range. It has also proved useful in fitting the
low-energy behaviour of NN scattering~\cite{orvk,md3,pkmr,sf}. However in
systems with unnaturally large scattering lengths, such an approach is not
systematic~\cite{md5}. It is thus unclear what advantage the parametrisation
(\ref{eq:potreg}) has over, say, a sum of Yukawa terms~\cite{reid}.

DR with minimal subtraction does lead to a consistent power
counting~\cite{ksw}. However in this scheme the divergent loop integrals are
all set to zero, and so the $K$-matrix is equal to the potential. As a result,
the range of validity of the momentum expansion is controlled by the
scattering length and, in systems with an unnaturally large scattering length,
this range is too small to be of practical use~\cite{ksw,lm}.

This leaves DR with power divergence subtraction~\cite{ksw2} as the only hope
for a renormalisation scheme that is both useful and systematic. In the PDS
scheme, a modified power counting does exist. This is based on taking the
renormalisation scale $\mu$ to be of the same order as the momenta of interest.
The potential and the scattering amplitude both contain terms with all integer
powers of the low scale $Q$ ($p$ or $\mu$) starting at order ${\cal
O}(Q^{-1})$. This is to be contrasted with Weinberg's counting~\cite{wein2},
which would be applicable to systems with a natural scattering length. In
that, the potential and $K$-matrix contain only even powers of $p$ starting at
order ${\cal O}(p^0)$\footnote{The $T$-matrix does contain odd powers of $p$,
but only because of the unitarity term $iM p/4\pi$ in its denominator.}.

In the PDS scheme each loop integral contributes one power of $p$ or $\mu$ 
beyond those associated with the vertices in the loop. This is true even if,
as suggested above, the scheme is modified to include subtraction terms for all
power-law divergences. Thus the power counting of Ref.~\cite{ksw2} is not
spoiled by keeping higher than linear divergences. One also sees that all
iterations of the leading term in the potential are of order ${\cal
O}(Q^{-1})$, and so should be summed up nonperturbatively. In contrast, higher
order terms in the short-range potential can be treated as perturbations (as
can pion-exchange if pions are included explicitly in the low-energy
theory~\cite{ksw2,mjs}).

\section{Renormalisation group ideas}

In sections 4 and 5 we have seen that power counting is possible in theories
with a natural scattering length, and that the method of regularisation does
not affect this conclusion. However, for systems with unnaturally large
scattering lengths, cut-off regularisation and DR with PDS lead to quite
different results. Although not the one suggested by Weinberg~\cite{wein1}, a
systematic power counting does emerge in the PDS scheme. In contrast no such
organisation of the theory is possible when a finite cut-off is used.

If the scale $\mu$ introduced by PDS is regarded as a resolution scale, then
one can understand why the PDS scattering equations have a rather similar form
to those obtained with a cut-off. However, the very different behaviour of the
potential in these regularisation schemes is then all the more surprising.
As described in sections 4 and 6, the origin of this difference lies in the
UV finite pieces of the cut-off loop integrals (\ref{eq:expand}) that are not
present in the PDS scheme.\footnote{For more on this, see D. Phillips' 
contribution to these proceedings~\cite{drp}.}

In this section, I discuss low-energy NN scattering from the viewpoint of the
renormalisation group (RG)~\cite{wilson,bt,morris}. These ideas were developed
following discussions at the workshop and may eventually help to clarify the
differences between the regularisation schemes. In an RG treatment, one
demands that the effective theory continue to reproduce physical quantities as
the the cut-off scale is lowered and so more and more short-distance physics
is ``integrated out". For example, one can require that the cut-off theory
reproduce the scattering observables for all values of the cut-off by
differentiating the expressions for these observables with respect to
$\Lambda$ and setting the derivatives equal to zero. This leads to a set of
Wilson-style RG equations for the $\Lambda$ dependence of the coefficients
$C_{2n}$ in the potential. These equations are very similar to the
corresponding ones for the $\mu$ dependence of the coefficients in the PDS
scheme~\cite{ksw2}. Alternatively one can avoid writing down differential RG
equations by inverting the expressions relating the scattering observables to
the $C_{2n}$ and $\Lambda$. This approach, which the one followed by the
Maryland group in Refs.~\cite{md4,md5}, leads directly to the solutions of the
RG equations.

Viewed from this perspective, it looks unnatural to try to take the cut-off 
to infinity as in Refs.~\cite{md4,md5,rbm}, in which case the breakdown of
power counting and related problems found in this limit do not rule out the
possibility of systematic low-energy effective theory. For example a similar
breakdown is expected in mesonic ChPT with a large cut-off (even if the
cut-off is imposed in a way that does not violate the symmetries of the
theory). Instead one needs to examine the behaviour of the theory as the
cut-off is lowered. If the couplings (suitably scaled by powers of $\Lambda$)
tend to definite values as $\Lambda\rightarrow 0$, then this IR fixed point
corresponds to a well-defined effective theory.

In the case of scattering by short-range potentials with a natural scattering
length, the rescaled couplings $\hat C_{0,2}$ defined in (\ref{eq:dless})
tend to the fixed point $\hat C_0=\hat C_2=0$ as $\Lambda\rightarrow 0$.
If higher-order terms are included in the potential then one finds that the
fixed point is the trivial one, $\hat C_{2n}=0$ for all $n$. The corresponding
effective theory is the one found using either a finite cut-off, where loop
corrections can be treated perturbatively for small $\Lambda$, or DR with
minimal subtraction, where loop corrections are ignored. As we have seen above,
a systematic power counting is possible in a perturbative low-energy theory of
this type.

In contrast PDS aims to treat systems with unnaturally large scattering
lengths. Strictly speaking, as $\Lambda$ tends to zero such a theory eventually
approaches a fixed point of the type discussed. However, in the region where
$\Lambda$ is much larger that $1/a$ but small compared with $1/r_e$, we can
neglect corrections of order $1/\Lambda a$ and consider the approach to a
fixed point (if one exists) that corresponds to a theory with infinite
scattering length. Such a ``quasi-fixed point" is central to the power counting
found in the PDS scheme~\cite{ksw2}. Let me first consider such systems with a
cut-off. In this case, one finds from (\ref{eq:erefc}) that to get a finite
effective range, both $\hat C_0$ and $\hat C_2$ must tend to finite values as
$\Lambda\rightarrow 0$. However there is no power counting: the value of $\hat
C_0$ at the fixed point changes by a finite amount when $\hat C_2$ is
included. Similarly both of these coefficients will change if higher-order
terms are included in the potential.

The results obtained with a cut-off imply that, if it exists, the fixed point
for a system with infinite scattering length is a nonperturbative one; the
corresponding values of the $\hat C_{2n}$ can only be determined if one
includes terms to all orders in the expansion of the potential. The
nonperturbative nature of this fixed point may mean that it is not accessible
using DR, unlike the perturbative point found for natural systems. Certainly
the PDS scheme shows a very different fixed-point behaviour. In this case, all
the $C_{2n}$ for $n>0$ tend to zero and only $\hat C_0$ tends to a finite
value. As already mentioned, this behaviour is different from that found using
a cut-off because DR does not pick up the terms in the loop integrals
(\ref{eq:expand}) that are proportional to inverse powers of $\Lambda$ and so
become increasingly important as $\Lambda\rightarrow 0$. These terms include 
the one involving $B_1$ in (\ref{eq:erefc}), which has the effect of driving
$\hat C_2$ to a finite value in this limit, along with similar ones at
higher-orders in the expansion.

By omitting the IR divergent terms from the loop diagrams, the PDS scheme
leads to a set of RG equations with a quite different fixed point behaviour
 from those obtained with a cut-off. It is therefore important to understand 
the nature of these terms~\cite{drp}. If they are not artefacts of cut-off
regularisation, then the failure to retain these terms in the RG equations
suggests that the apparent fixed point of PDS does not really correspond to a
well-defined low-energy effective theory. If this is the case, a systematic
power counting would only be possible for systems with natural scattering
lengths. In the case of interest, $s$-wave NN scattering, power counting would
indeed be powerless.

\section*{Acknowledgments}

I am grateful to J. McGovern and K. Richardson for their collaboration on
Ref.~\cite{rbm} and to them, D. Phillips and U. van Kolck for helpful
discussions. I also wish to thank the organisers of the Caltech/INT workshop
for a very stimulating meeting, M. Wise and D. Kaplan for their pertinent
questions during my talk, and D. Phillips for providing what I believe is
the key to answering them. This work was supported by the EPSRC and PPARC.

\end{document}